\documentclass[twocolumn,aps,prd,groupedaddress,nofootinbib,preprintnumbers,showpacs,superscriptaddress]{revtex4}
\usepackage[T1]{fontenc} 
\usepackage{graphicx}
\usepackage{amsmath}
\usepackage{bm}
\usepackage{slashed}
\usepackage{epstopdf}
\usepackage{epsfig}
\usepackage{verbatim}
\usepackage{mathrsfs}
\usepackage{bm}


\def\g0{\gamma_0}

\def\nn{\nonumber}
\def\beqa{\begin{eqnarray}}
\def\eeqa{\end{eqnarray}}
\def\beqn{\begin{eqnarray}}
\def\eeqn{\end{eqnarray}}
\def\beq{\begin{equation}}
\def\eeq{\end{equation}}

\def\I4{I_4}

\begin{document}

\preprint{DESY~19-098\hspace{13cm}ISSN~0418-9833}
\preprint{December 2019\hspace{15cm}}


\title{Inclusive $J/\psi$ and $\eta_c$ production in $\Upsilon$ decay at
$\mathcal{O}(\alpha_s^5)$ in nonrelativistic QCD factorization}

\author{Zhi-Guo He}
\email{zhiguo.he@desy.de}
\affiliation{{II.} Institut f\"ur Theoretische Physik, Universit\"at Hamburg,
Luruper Chaussee 149, 22761 Hamburg, Germany}

\author{Bernd A. Kniehl}
\email{kniehl@desy.de}
\affiliation{{II.} Institut f\"ur Theoretische Physik, Universit\"at Hamburg,
Luruper Chaussee 149, 22761 Hamburg, Germany}

\author{Xiang-Peng Wang}
\email{xiangpeng.wang@anl.gov}
\affiliation{{II.} Institut f\"ur Theoretische Physik, Universit\"at Hamburg,
Luruper Chaussee 149, 22761 Hamburg, Germany}
\affiliation{High Energy Physics Division, Argonne National Laboratory,
Argonne, Illinois 60439, USA}

\date{\today}

\begin{abstract}
We study $J/\psi$ and $\eta_c$ inclusive production in $\Upsilon$ decay within
the framework of nonrelativistic-QCD (NRQCD) factorization.
In the latter case, for which no experimental data exist so far, we also
include the $h_c$ feed-down contribution.
We calculate the short-distance coefficients completely through
$\mathcal{O}(\alpha_s^5)$.
The NRQCD predictions for the branching fraction
$\mathcal{B}(\Upsilon\to J/\psi+X)$ via direct production,
evaluated with different sets of long-distance matrix elements (LDMEs), all
agree with the experimental data in a reasonable range of renormalization scale.
Using $\eta_c$ and $h_c$ LDMEs obtained from $J/\psi$ and $\chi_c$ ones via
heavy-quark spin symmetry, we find that the bulk of
$\mathcal{B}(\Upsilon\to\eta_c+X)$ via prompt production arises from the
$c\bar{c}({}^3\!S_1^{[8]})$ Fock state.
The experimental study of this decay process would, therefore, provide a
particularly clean probe of the color octet mechanism of heavy-quarkonium
production.
\end{abstract}

\pacs{12.38.Bx, 12.39.St, 13.25.Gv, 14.40.Pq}

\maketitle

\section{Introduction}

Heavy-quarkonium production serves as an ideal laboratory to study the
interplay of perturbative and nonperturbative phenomena of QCD thanks to the
hierarchy of energy scales $m_Q v_Q^2\ll m_Qv_Q\ll m_Q$ characterizing kinetic
energy, momentum, and mass, where $m_Q$ is the mass of the heavy quark $Q$
and $v_Q$ is its relative velocity in the rest frame of the heavy quarkonium.
The effective quantum field theory of nonrelativistic QCD (NRQCD)
\cite{Caswell:1985ui} endowed with the factorization conjecture of
Ref.~\cite{Bodwin:1994jh} is the standard theoretical approach to study
quarkonium production and  decay.
This conjecture states that the theoretical predictions can be separated into
process-dependent short-distance coefficients (SDCs) calculated perturbatively
as expansions in the strong-coupling constant $\alpha_s$ and supposedly
universal long-distance matrix  elements (LDMEs), scaling with definite powers
of $v_Q$ \cite{Lepage:1992tx}.
In this way, the theoretical calculations are organized as double expansion in
$\alpha_s $ and $v_Q$.

During the past quarter of a century, the NRQCD factorization approach has
been very successful in describing both heavy-quarkonium production and decay;
see Refs.~\cite{Brambilla:2010cs,Brambilla:2014jmp,Lansberg:2019adr} for
reviews.
However, there are still open problems in charmonium production, in particular
for the $J/\psi$ meson.
Prompt $J/\psi$ production has been studied in various scenarios both
experimentally and theoretically.
Specifically, the SDCs are known at next-to-leading order (NLO) in $\alpha_s$
for the yield \cite{Zhang:2009ym,Ma:2008gq} and polarization \cite{Gong:2009kp}
in $e^+e^-$ annihilation, the yield in two-photon collisions
\cite{Klasen:2004tz,Klasen:2004az,Butenschoen:2011yh}, the yield
\cite{Butenschoen:2009zy} and polarization \cite{Butenschoen:2011ks} in
photoproduction, the yield \cite{Ma:2010yw,Butenschoen:2010rq} and polarization
\cite{Butenschoen:2012px,Chao:2012iv,Gong:2012ug,Shao:2014yta} in
hadroproduction, etc.
Different sets of LDMEs were obtained by fitting experimental data adopting
different strategies.
Unfortunately, none of them can explain all the experimental measurements,
which challenges the universality of the NRQCD LDMEs.
Moreover, it has been found \cite{Butenschoen:2014dra} that all the LDME sets
determined from $J/\psi$ production data, upon conversion to $\eta_c$ LDMEs via
heavy-quark spin symmetry, result in NLO predictions that overshoot the
$\eta_c$ hadroproduction data, taken by the LHCb Collaboration at the CERN
Large Hadron Collider (LHC) \cite{Aaij:2014bga}.
To shed more light on this notorious $J/\psi$ puzzle, it is useful to consider
yet further production modes.
This provides an excellent motivation to study $J/\psi$ and $\eta_c$ inclusive
production in $\Upsilon$ decays, which is the goal of this paper.
This complements our recent study of $\chi_{cJ}$ inclusive production in
$\Upsilon$ decays \cite{He:2018dho}.
To facilitate the comparison with experimental data, we will focus here on
direct $J/\psi$ production, while we will consider prompt $\eta_c$ production,
by also including the feed-down from $h_c$ mesons.

On the experimental side, the $\Upsilon\to J/\psi+X$ decay process was first
observed by the CLEO Collaboration at the Cornell Electron Storage Ring CESR
about three decades ago \cite{Fulton:1988ug}.
Several independent measurements
\cite{Maschmann:1989ai,Albrecht:1992ap,Briere:2004ug} followed.
The latest one was carried out by the Belle Collaboration at the KEK $B$
factory KEKB \cite{Shen:2016yzg}.
The present world average of the branching fraction for prompt production is
\cite{Tanabashi:2018oca}
\begin{equation}
\mathcal{B}_\text{prompt}(\Upsilon\to J/\psi+X)=(5.4\pm0.4) \times 10^{-4}.
\end{equation}
As for the $\Upsilon\to\eta_c+X$ decay, there is no experimental data yet.
With the large amount of data to be accumulated by the Belle II Collaboration
at the SuperKEKB accelerator, there might be a chance to study this process, and
NRQCD predictions for its branching fraction will be needed.

On the theoretical side, the $\Upsilon\to J/\psi+X$ process was proposed as a
rich gluon environment to study the color octet (CO) mechanism of NRQCD
factorization in Refs.~\cite{Cheung:1996mh,Napsuciale:1997bz}, where some of
the CO channels, including 
$b\bar{b}({}^3\!S_1^{[1]})\to c\bar{c}({}^3\!S_1^{[8]})+gg$ at
$\mathcal{O}(\alpha_s^4)$ and 
$b\bar{b}({}^3\!S_1^{[1]})\to c\bar{c}({}^1\!S_0^{[8]},{}^3\!P_J^{[8]})+g$ at
$\mathcal{O}(\alpha_s^5)$, were taken into account.
Later, the color singlet (CS) processes, including
$b\bar{b}({}^3\!S_1^{[1]})\to c\bar{c}({}^3\!S_1^{[1]})+c\bar{c}g$ at
$\mathcal{O}(\alpha_s^5)$,
$b\bar{b}({}^3\!S_1^{[1]})\to c\bar{c}({}^3\!S_1^{[1]})+gg(gggg)$ at
$\mathcal{O}(\alpha_s^6)$, and some  interesting QED contribution were also
calculated \cite{He:2009by,He:2010cb}.
It was found that the CS contribution itself is about 3.8 times smaller than
the experimental data, which suggests the potential need for a large CO
contribution.

In the previous studies
\cite{Cheung:1996mh,Napsuciale:1997bz,He:2009by,He:2010cb}, only the
leading-order (LO) $\mathcal{O}(\alpha_s^4)$ contribution and some part of the
$\mathcal{O}(\alpha_s^5)$ channels were considered and pre-LHC LDME sets were
used.
In this work, we will obtain the complete SDC results through
$\mathcal{O}(\alpha_s^5)$, both for $J/\psi$ and $\eta_c$ production, and
perform a numerical analysis with up-to-date LDMEs sets. 
On top of direct production, prompt production also includes the feed-down from
heavier charmonia, namely from $\chi_{cJ}$ and $\psi^\prime$ mesons in the
$J/\psi$ case and from $h_c$ mesons in the $\eta_c$ case.
The direct $J/\psi$ results obtained here readily carry over to $\psi^\prime$
feed-down, while the $\chi_{cJ}$ results may be found in Ref.~\cite{He:2018dho}.
The $h_c$ feed-down results will also be derived here. 

This paper is organized as follows.
In Sec.~\ref{sec:two}, we describe our analytical calculations.
In Sec.~\ref{sec:three}, we present and interpret our numerical results.
Section~\ref{sec:four} contains our conclusions.
In the Appendix, 
we list the contributing SDCs at $\mathcal{O}(\alpha_s^3)$.

\section{Analytical calculations}
\label{sec:two}

According to NRQCD factorization \cite{Bodwin:1994jh}, the partial decay width
of $\Upsilon\to H+X$ with $H=J/\psi,\eta_c,h_c$ is given by 
\begin{equation}
  \Gamma (\Upsilon \to H + X)=\sum_{m,n}\hat{\Gamma}_{mn}
  \langle \Upsilon | \mathcal{O}(m) |\Upsilon\rangle
  \langle \mathcal{O}^{H}(n) \rangle,
\label{eq:decay}
\end{equation}
where $\hat{\Gamma}_{mn}=\hat{\Gamma}(b\bar{b}(m)\to c\bar{c}(n)+X)$ is the SDC
and $\langle \Upsilon | \mathcal{O}(m) |\Upsilon\rangle$ and
$\langle \mathcal{O}^{H}(n) \rangle$ are the LDMEs of $\Upsilon$ decay and $H$
production, respectively.

For $\Upsilon$, the CO contribution is so small~\cite{Cheung:1996mh} that we
only include the CS case of $m={}^3\!S_1^{[1]}$.
In fact, by the velocity scaling rules \cite{Lepage:1992tx}, the leading CO
LDMEs of $\Upsilon$ decay are parametrically suppressed by a factor of
$v_b^4\approx 1\%$ relative to the CS one,
$\langle\Upsilon|\mathcal{O}({}^3S_1^{[1]})|\Upsilon\rangle$.
In Ref.~\cite{Bodwin:2005gg}, the suppression factors for the $S$ wave CO
LDMEs of $\Upsilon$ decay were quantitatively determined in the framework of
lattice NRQCD using two different schemes of implementing heavy-quark Green's
functions, the hybrid and nrqcd schemes.
The largest values were obtained in the hybrid scheme, namely,
$\langle\Upsilon|\mathcal{O}({}^1S_0^{[8]})|\Upsilon\rangle
/\langle\Upsilon|\mathcal{O}({}^3S_1^{[1]})|\Upsilon\rangle
=2.414(3)\times10^{-3}$
and
$\langle\Upsilon|\mathcal{O}({}^3S_1^{[8]})|\Upsilon\rangle
/\langle\Upsilon|\mathcal{O}({}^3S_1^{[1]})|\Upsilon\rangle
=8.1(6)\times10^{-5}$;
see Table~I in Ref.~\cite{Bodwin:2005gg}.
Similar analyses for the $P$ wave LDME
$\langle\Upsilon|\mathcal{O}({}^3P_0^{[8]})|\Upsilon\rangle$ are not yet
available, but its suppression factor is expected to be in the same ballpark,
of $\mathcal{O}(10^{-3})$ or below.
On the other hand, SDCs involving CO $b\bar{b}$ Fock states $m$ already start
at $\mathcal{O}(\alpha_s^3)$.
However, using $\alpha_s(m_b)\approx0.26$, we have
$\alpha_s^2\approx 0.068\gg10^{-3}$, so that the contributions to
Eq.~(\ref{eq:decay}) from the leading CO $b\bar{b}$ Fock states,
$m={}^1\!S_0^{[8]},{}^3\!S_1^{[8]},{}^3\!P_J^{[8]}$, are expected to be safely
negligible against the $\mathcal{O}(\alpha_s^5)$ contributions with
$m={}^3\!S_1^{[1]}$, which we still include here.
For future use, we list the $\mathcal{O}(\alpha_s^3)$ SDCs $\hat{\Gamma}_{mn}$
for all the decay processes considered here and in Ref.~\cite{He:2018dho},
$\Upsilon\to H+X$ with $H=J/\psi,\eta_c,\chi_{cJ},h_c$, in the Appendix.

On the other hand, for $H$, we include the CO contributions of LO in $v_c^2$
besides the CS contributions, {\it i.e.}\ we have
$n={}^3\!S_1^{[1,8]},{}^1\!S_0^{[8]},{}^3\!P_J^{[8]}$ for $J/\psi$,
$n={}^1\!S_0^{[1,8]},{}^3\!S_1^{[8]},{}^1\!P_1^{[8]}$ for $\eta_c$, and
$n={}^1\!P_1^{[1]},{}^1\!S_0^{[8]}$ for $h_c$.

In our computation, we generate all the Feynman amplitudes using the FeynArts 
package~\cite{Hahn:2000kx}.
The Dirac and color operations are performed with the help of the
FeynCalc~\cite{Mertig:1990an} and FORM~\cite{Kuipers:2012rf} packages.
We use the Mathematica package {\$}Apart~\cite{Feng:2012iq} to decompose
linearly dependent propagators in the loop integrals to irreducible ones,
which we further reduce to master scalar integrals using the
FIRE package \cite{Smirnov:2008iw}.
We then evaluate the master integrals numerically using the C$++$ package
QCDLOOP~\cite{Carrazza:2016gav}.
Finally, we perform the phase space integrations numerically using the
CUBA~\cite{Hahn:2004fe} library.

\begin{figure}
\begin{center}
\includegraphics[scale=0.3]{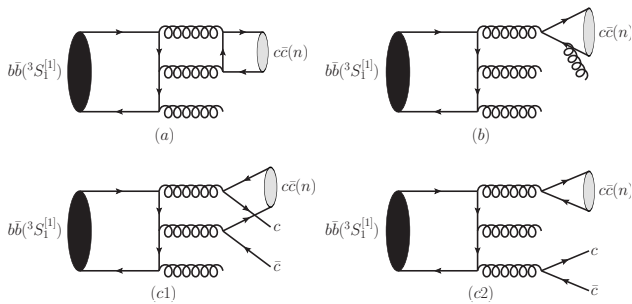}
\caption{Representative Feynman diagrams for subprocesses
  $b\bar{b}({}^3\!S_1^{[1]})\to c\bar{c}(n)+X$, with (a) $X=g$, (b) $X=ggg$, and
  (c) $X=c\bar{c}g$.
  The possible $c\bar{c}$ Fock states include $n={}^1\!S_0^{[8]},{}^3\!P_J^{[8]}$
  in group (a), $n={}^1\!S_0^{[1,8]},{}^1\!P_1^{[8]},{}^3\!P_J^{[8]}$ in group (b),
  $n={}^1\!S_0^{[1,8]},{}^3\!S_1^{[1,8]},{}^1\!P_1^{[1,8]},{}^3\!P_J^{[8]}$ in
  subgroup (c1), and $n={}^3\!S_1^{[8]}$ in subgroup (c2).}
\label{Feyndiag}
\end{center}
\end{figure}

The $\mathcal{O}(\alpha_s^4)$ LO contribution only includes the
$b\bar{b}({}^3\!S_1^{[1]})\to c\bar{c}({}^3\!S_1^{[8]})+gg$ subprocess.
Its NLO QCD corrections, of $\mathcal{O}(\alpha_s^5)$, were calculated in our
previous work~\cite{He:2018dho}, where details may be found.
Here, we only present numerical results.
At $\mathcal{O}(\alpha_s^5)$, also a number of new partonic production channels
open up.
We derive here the SDCs of all of them.
In the cases where results already exist in the literature, we compare and find
agreement. 
For the sake of a systematical discussion, we divide the contributing
partonic subprocesses into three groups according to the parton content of the
system $X$ in the final state: (a) $X=g$, (b) $X=ggg$, and (c) $X=c\bar{c}g$.
Representative Feynman diagrams for each group are shown in 
Fig.~\ref{Feyndiag}.      

In group (a), the only possible $c\bar{c}$ Fock states are
$n={}^1\!S_0^{[8]},{}^3\!P_J^{[8]}$, and there are 6 Feynman diagrams, a typical
one of which is shown in panel (a) of Fig.~\ref{Feyndiag}. 
The SDCs of these partonic subprocesses were first calculated in
Ref.~\cite{Napsuciale:1997bz}.
We evaluate the one-loop amplitude of
$b\bar{b}({}^3\!S_1^{[1]})\to c\bar{c}({}^1\!S_0^{[8]})+g$ applying the helicity 
projector approach \cite{Korner:1982vg} and the one of
$b\bar{b}({}^3\!S_1^{[1]})\to c\bar{c}({}^3\!P_J^{[8]})+g$ as the virtual
corrections in Ref.~\cite{He:2018dho}.
We find agreement with Ref.~\cite{Napsuciale:1997bz}.

\begin{figure}
\begin{center}
\includegraphics[width=0.9\linewidth]{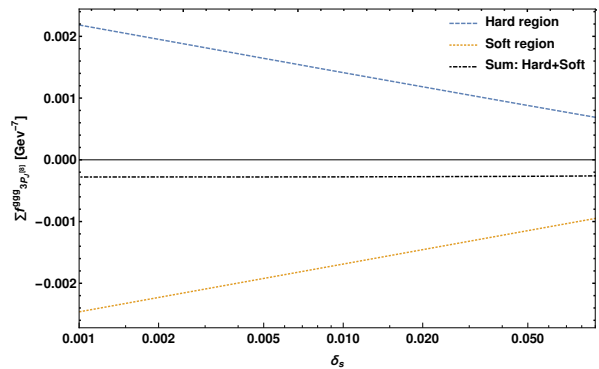}
\caption{Numerical dependence on the phase space slicing parameter $\delta_s$
of the soft (dotted line) and hard (dashed line) parts of the SDC of
$b\bar{b}({}^3\!S_1^{1})\to c\bar{c}({}^3\!P_J^{[8]}) + ggg$ as well as their
superposition (dot-dashed line) upon subtraction of the infrared divergence at
NRQCD factorization scale $\mu_{\Lambda}=m_c$.}
\label{DeltasPCO}
\end{center}
\end{figure}

In group (b), the possible $c\bar{c}$ Fock states are
$n={}^1\!S_0^{[1,8]},{}^1\!P_1^{[8]},{}^3\!P_J^{[8]}$, and there are 36 Feynman
diagrams, a representative one of which is displayed in panel (b) of
Fig.~\ref{Feyndiag}.
None of these partonic subprocesses have been studied before.
Except for $n={}^3\!P_J^{[8]}$, all the SDCs are infrared finite and
straightforwardly calculated.
The infrared divergence in the SDC of
$b\bar{b}({}^3\!S_1^{[1]})\to c\bar{c}({}^3\!P_J^{[8]}) + ggg$ can be extracted by
means of the phase space slicing method \cite{Harris:2001sx}, as explained in
detail for the CS case of ${}^3\!P_J^{[1]}$ in Ref.~\cite{He:2018dho}.
In NRQCD factorization, this infrared divergence is absorbed into the NLO
corrections to the CO LDME  $\langle\mathcal{O}^{J/\psi}({}^3\!S_1^{[8]})\rangle$
\cite{Bodwin:1994jh,Butenschon:2009zza},
\beqn
\lefteqn{\langle{\cal O}^{J/\psi}({}^3\!S_1^{[8]})\rangle_{\text{Born}}
  = \langle{\cal O}^{J/\psi}({}^3\!S_1^{[8]})\rangle_{\text{ren}}}\nn\\
&&{}+\frac{2\alpha_s}{3\pi m_c^2}
\left(\frac{4\pi\mu^2}{\mu^2_\Lambda}e^{-\gamma_E}\right)^\epsilon
\frac{1}{\epsilon_{\text{IR}}}\sum_J\left[\frac{C_F}{C_A}
  \langle{\cal O}^{J/\psi}({}^3\!P_J^{[1]})\rangle_{\text{Born}}\right.\nn\\
  &&{}+\left.\left(\frac{C_A}{2}-\frac{2}{C_A}\right)
  \langle{\cal O}^{J/\psi}({}^3\!P_J^{[8]})\rangle_{\text{Born}}\right],
\eeqn 
where $\langle{\cal O}^{J/\psi}({}^3\!S_1^{[8]})\rangle_{\text{ren}}$ is the
renormalized LDME, $\mu$ and $\mu_\Lambda$ are the QCD and NRQCD factorization
scales, respectively, $\epsilon=2-d/2$ in $d$ space-time dimensions, and
$C_F=4/3$ and $C_A=3$ are color factors.
We verified numerically that the finite piece left over after removing the
infrared divergence does not depend on the slicing parameter $\delta_s$, which
is illustrated in Fig.~\ref{DeltasPCO}. 

In group (c), all the relevant $c\bar{c}$ Fock states
$n={}^1\!S_0^{[1,8]},{}^3\!S_1^{[1,8]},{}^1\!P_1^{[1,8]},{}^3\!P_J^{[8]}$ appear.
There are generally 6 Feynman diagrams similar to the one shown in
panel (c1) of Fig.~\ref{Feyndiag}, with the exception of $n={}^3\!S_1^{[8]}$
for which there are 6 additional ones of fragmentation type as in
panel (c2).
The SDC of $b\bar{b}({}^3\!S_1^{[1]})\to c\bar{c}({}^3\!S_1^{[1]}) + c\bar{c}g$
was first considered in Ref.~\cite{Li:1999ar}, and the correct result was first
presented in Ref.~\cite{He:2009by}. 
All the other partonic subprocesses are studied here for the first time.
Their SDCs are all infrared finite and can be calculated straightforwardly.

To be able to investigate the relative weight of each partonic subprocess, we
decompose $\hat{\Gamma}_{{}^3\!S_1^{[1]},n}$ for each $n$ according to the possible
final-state systems $X$,
\begin{equation}
  \hat{\Gamma}_{{}^3\!S_1^{[1]},n}=\sum_{X}\hat{\Gamma}_{n}^X,
\end{equation}
where $X=gg,ggg,c\bar{c}g$ for $n={}^3\!S_1^{[8]}$,
$X=g,ggg,c\bar{c}g$ for $n={}^1\!S_0^{[8]},{}^3\!P_J^{[8]}$,
$X=ggg,c\bar{c}g$ for $n={}^1\!S_0^{[1]},{}^1\!P_1^{[8]}$, and
$X=c\bar{c}g$ for $n={}^3\!S_1^{[1]},{}^1\!P_1^{[1]}$.
Furthermore, we factor out the strong-coupling constant $\alpha_s(\mu)$ and
render the coefficients dimensionless by writing
\begin{eqnarray}\label{eq:fs}
  \hat{\Gamma}_{{}^3\!S_1^{[8]}}^{gg}&=&[f_{{}^3\!S_1^{[8]}}^{gg,\text{LO}}
    +f_{{}^3\!S_1^{[8]}}^{gg,\text{corr}}(\mu)\alpha_s(\mu)]\alpha_s^4(\mu)
  \text{GeV}^{-5},
  \nonumber\\
  \hat{\Gamma}_{n}^{X}&=&f_{n}^{X}\alpha_s^{5}(\mu)\text{GeV}^{-k},
\end{eqnarray}
where $k=5$ for $n={}^3\!S_1^{[8]}$ with $X=c\bar{c}g$ and
$n={}^1\!S_0^{[1,8]},{}^3\!S_1^{[1]}$, and $k=7$ for
$n={}^1\!P_1^{[1,8]},{}^3\!P_J^{[8]}$.

\section{Numerical results}
\label{sec:three}

\begin{table*}[t!]
  \caption{Numerical results for the dimensionless coefficients $f_{n}^{X}$ in 
    Eq.~(\ref{eq:fs}) besides
$f_{{}^3\!S_1^{[8]}}^{gg,\text{corr}}=(4.85 +13.62\times\ln\frac{\mu}{m_b})\times10^{-4}$.}
\begin{ruledtabular}
\begin{tabular}{|c|c|c|c|c|c|c|c|c|c|c|c|c|c|}
$f_{{}^1\!S_0^{[1]}}^{ggg}$ &
$f_{{}^1\!S_0^{[1]}}^{c\bar{c}g}$ &
$f_{{}^3\!S_1^{[1]}}^{c\bar{c}g}$ &
$f_{{}^1\!P_1^{[1]}}^{c\bar{c}g}$ & 
$f_{{}^1\!S_0^{[8]}}^{g}$ & 
$f_{{}^1\!S_0^{[8]}}^{ggg}$ & 
$f_{{}^1\!S_0^{[8]}}^{c\bar{c}g}$ &
$f_{{}^3\!S_1^{[8]}}^{gg,\text{LO}}$ &
$f_{{}^3\!S_1^{[8]}}^{c\bar{c}g}$ &
$f_{{}^1\!P_1^{[8]}}^{ggg}$ &
$f_{{}^1\!P_1^{[8]}}^{c\bar{c}g}$ &
$f_{{}^3\!P_J^{[8]}}^{g}$ &
$f_{{}^3\!P_J^{[8]}}^{ggg}$ &
$f_{{}^3\!P_J^{[8]}}^{c\bar{c}g}$ \\
\hline
$1.89$ &
$1.07$ &
$1.32$ &
$1.5$ &
$1.15$ &
$1.50$ &
$2.14$ &
$2.38$ & 
$1.23$ &
$6.4$ &
$3.0$ &
$1.97$ &
$-2.78$ &  
$1.3$ \\
$\times10^{-6}$ &
$\times10^{-6}$ &
$\times10^{-6}$ &
$\times10^{-7}$ & 
$\times10^{-4}$ & 
$\times10^{-5}$ & 
$\times10^{-6}$ &
$\times10^{-4}$ &
$\times10^{-5}$ &
$\times10^{-7}$ &
$\times10^{-7}$ &
$\times10^{-4}$ &
$\times10^{-4}$ &
$\times10^{-6}$ \\
\end{tabular}
\end{ruledtabular}
\label{SDCs}
\end{table*}

In our numerical analysis, we evaluate $\alpha_s(\mu)$ with $n_f=3$ massless
quark flavors and $\Lambda_\text{QCD}=249~(389)~\text{MeV}$ at LO (NLO), set
$m_c=1.5~\text{GeV}$, $m_b=4.75~\text{GeV}$, and choose $\mu_{\Lambda}=m_c$.
The resulting values of $f_n^{X}$ are presented in Table~\ref{SDCs}.
The only instance where the LO formula for $\alpha_s(\mu)$ is used is in the
LO evaluation of $\hat{\Gamma}_{{}^3\!S_1^{[8]}}^{gg}$ in Eq.~(\ref{eq:fs}) with
$f_{{}^3\!S_1^{[8]}}^{gg,\text{corr}}(\mu)$ put to zero.

Given the large powers of $\alpha_s(\mu)$ appearing in Eq.~(\ref{eq:fs}), we
are faced with considerable $\mu$ dependencies, so that scale optimization
appears appropriate.
As in Ref.~\cite{He:2018dho}, we adopt the principle of fastest apparent
convergence (FAC) \cite{Grunberg:1980ja} to determine the default value of
$\mu$, by requiring that the LO and NLO evaluations of
$\hat{\Gamma}_{{}^3\!S_1^{[8]}}^{gg}$ as described above coincide.
This yields $\mu_{\text{FAC}}=6.2~\text{GeV}$.
It would then be natural to explore the $\mu$ dependence in the range
$\mu_{\text{FAC}}/2<\mu<2\mu_{\text{FAC}}$.
However, the NLO result for $\mathcal{B}(\Upsilon\to \chi_{c0} +X)$ is known
to be negative for $\mu<3.7~\text{GeV}$ \cite{He:2018dho}.
We thus consider the reduced $\mu$ range $3.7~\text{GeV}<\mu<2\mu_{\text{FAC}}$
in the following.

When the $b\bar{b}$ pair is in a CS Fock state, it couples to at least three
gluons in the decays $\Upsilon\to H+X$ with $H=J/\psi,\eta_c,h_c$, as may be
seen in Fig.~\ref{Feyndiag}.
Therefore, the large theoretical uncertainty due to the freedom in the choices
of $\alpha_s$ and $m_b$ can be greatly reduced by normalizing
$\Gamma(\Upsilon\to H+X)$ with respect to $\Gamma(\Upsilon\to ggg)$
\cite{He:2009by,He:2010cb,He:2018dho}.
We thus write the branching fractions of interest as
\begin{equation}
  \mathcal{B}(\Upsilon\to H+X)
  =\frac{\Gamma(\Upsilon\to H+X)}{\Gamma(\Upsilon\to ggg)}
\mathcal{B}(\Upsilon\to ggg),
\label{eq:norm}
\end{equation}
where $\mathcal{B}(\Upsilon\to ggg)=81.7\%$ \cite{Tanabashi:2018oca}.
Through $\mathcal{O}(\alpha_s^4)$, we have \cite{Mackenzie:1981sf}
\begin{eqnarray}
  \lefteqn{\Gamma(\Upsilon\to ggg)=\frac{20\alpha_s^{3}(\mu)}{243m_b^2}(\pi^2-9)
\langle\Upsilon|\mathcal{O}({}^3\!S_1^{[1]})|\Upsilon\rangle}
\nonumber\\
&&{}\times\left\{1+\frac{\alpha_s(\mu)}{\pi}\left[-19.4
  +\frac{3\beta_0}{2}\left(1.161+\ln{\frac{\mu}{m_b}}\right)\right]\right\},
\quad
\end{eqnarray}
where $\beta_0=11-2n_f/3$ with $n_f=4$.
As an additional benefit, the theoretical predictions do not depend on the CS
LDME $\langle\Upsilon|\mathcal{O}({}^3\!S_1^{[1]})|\Upsilon\rangle$ anymore. 

\subsection{$\Upsilon\to J/\psi+X$}

\begingroup
\squeezetable
\begin{table*}
  \caption{$J/\psi$ LDME sets from
    Refs.~\cite{Butenschoen:2011yh,Butenschoen:2012qh,Chao:2012iv,Gong:2012ug,%
      Bodwin:2015iua,Han:2014jya}.
    Set~3 of Ref.~\cite{Chao:2012iv} only gave an upper bound on
    $\langle {\cal O}^{J/\psi}({}^1\!S_0^{[8]})\rangle$ and is replaced here by
    its update in Ref.~\cite{Han:2014jya}.}
\begin{ruledtabular}
  \begin{tabular}{|c|c|c|c|c|c|c|}
 $J/\psi$ LDME set
 & Butensch\"on {\it et al.}{}
 & Gong {\it et al.}{}
 & Bodwin {\it et al.}{}
 & Chao {\it et al.}{}
 & Chao {\it et al.}{}
 & Chao {\it et al.}{} \\
 & \cite{Butenschoen:2011yh,Butenschoen:2012qh}  
 & \cite{Gong:2012ug}
 & \cite{Bodwin:2015iua}
 & Default set \cite{Chao:2012iv}
 & Set 2 \cite{Chao:2012iv}
 & Set 3 \cite{Han:2014jya} \\
 \hline
 $\langle {\cal O}^{J/\psi}({}^3\!S_1^{[1]}) \rangle /\text{GeV}^3 $
 & $1.32$
 & $1.16$ 
 & $1.32$ 
 & $1.16$ 
 & $1.16$
 & $1.16$ \\
 $\langle {\cal O}^{J/\psi}({}^1\!S_0^{[8]}) \rangle /\text{GeV}^3$
 & $0.0304\pm0.0035$  
 & $0.097\pm0.009$
 & $0.110\pm0.014$
 & $0.089\pm0.0098$
 & $0$
 & $0.0146\pm0.0020$ \\
 $\langle {\cal O}^{J/\psi}({}^3\!S_1^{[8]}) \rangle /\text{GeV}^3$
 & $0.00168\pm0.00046$
 & $-0.0046\pm0.0013$
 & $-0.00713\pm0.00364$
 & $0.0030\pm0.0012$
 & $0.014$
 & $0.00903\pm0.00275$ \\
 $\langle {\cal O}^{J/\psi}({}^3\!P_0^{[8]}) \rangle /\text{GeV}^5$
 & $-0.00908\pm0.00161$
 & $-0.0214\pm0.0056$
 & $-0.00702\pm0.00340$
 & $0.0126\pm0.0047$
 & $0.054$
 & $0.0343\pm0.0110$ \\
 \end{tabular}
\end{ruledtabular}
\label{LDME} 
\end{table*} 
\endgroup

We first discuss $\Upsilon\to J/\psi+X$ via direct production.
From Table~\ref{SDCs}, we observe that all $c\bar{c}$ Fock states
${}^3\!S_1^{[1]}$, ${}^3\!S_1^{[8]}$, ${}^1\!S_0^{[8]}$, ${}^3\!P_J^{[8]}$ contribute 
significantly as long as the values of the corresponding LDMEs are not too
small.
Incidentally, there is a strong cancellation between the
$c\bar{c}({}^3\!P_J^{[8]})+g$ and $c\bar{c}({}^3\!P_J^{[8]})+ggg$ channels, the
latter one being studied here for the first time. 
Several $J/\psi$ LDME sets have been determined, both at LO and NLO, by
fitting to different sets of $J/\psi$ yield and polarization data.
There are significant differences between them, which reflects the potential
challenge to NRQCD factorization mentioned above.
We select here the ones most frequently used in the recent literature
\cite{Butenschoen:2011yh,Butenschoen:2012qh,Gong:2012ug,Bodwin:2015iua,%
Han:2014jya} and summarize their values in Table~\ref{LDME}.
Besides the contributions through $\mathcal{O}(\alpha_s^5)$ discussed above,
we will also include some other non-negligible contributions previously
studied, namely the pure CS contributions at $\mathcal{O}(\alpha_s^2\alpha^2)$
\cite{He:2009by} and $\mathcal{O}(\alpha_s^6)$ \cite{He:2010cb}.

\begin{table}
  \caption{Fitted $\mu_\text{fit}$ values for the $J/\psi$ LDME sets in
    Table~\ref{LDME}.}
  \label{tab:Jpsi}
\begin{ruledtabular}  
\begin{tabular}{|c|c|}
$J/\psi$ LDME set & $\mu_{\text{fit}}$ (GeV) \\
\hline
Butensch\"on {\it et al.}\ \cite{Butenschoen:2011yh,Butenschoen:2012qh} &
$3.8$ \\
\hline
Gong {\it et al.}\ \cite{Gong:2012ug} & $3.9$ \\
\hline
Bodwin {\it et al.}\ \cite{Bodwin:2015iua} & $3.7$ \\
\hline
Chao {\it et al.}\ (Default) \cite{Chao:2012iv} & $4.6$ \\
\hline
Chao {\it et al.}\ (Set 2) \cite{Chao:2012iv} & $5.1$ \\
\hline
Chao {\it et al.}\ (Set 3) \cite{Han:2014jya} & $4.2$ \\
\end{tabular}
\end{ruledtabular}  
\end{table}

\begin{figure}
\centering
\includegraphics[width=0.9\linewidth]{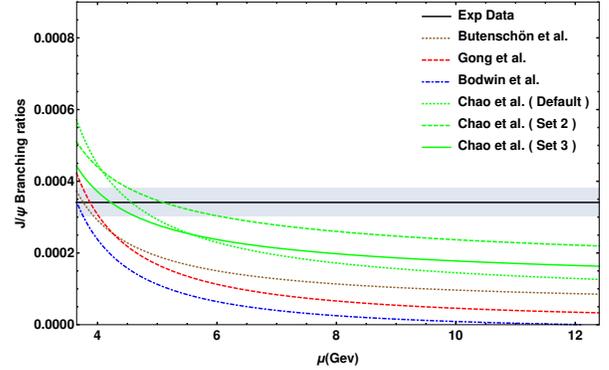}
\caption{$\mu$ dependencies of the NRQCD predictions for
  $\mathcal{B}_{\text{direct}}(\Upsilon\to J/\psi+X)$ based on the $J/\psi$ LDME
  sets in Table~\ref{LDME} compared with the experimental result in
  Eq.~(\ref{eq:pdg}).}
\label{JpsiResult}
\end{figure}

From Ref.~\cite{Tanabashi:2018oca}, we extract the result for direct $J/\psi$
production,
\begin{equation}
  \mathcal{B}_{\text{direct}}(\Upsilon\to J/\psi+X)=(3.46\pm 0.67)\times10^{-4},
  \label{eq:pdg}
\end{equation}
by subtracting the contributions due to the feed-down from the $\chi_{cJ}$ and
$\psi^\prime$ mesons.
In Fig.~\ref{JpsiResult}, we confront it with our NRQCD predictions evaluated
with the $J/\psi$ LDME sets listed in Table~\ref{LDME} as functions of $\mu$.
We observe from Fig.~\ref{JpsiResult} that, although the various predictions
appreciably differ in normalization and line shape, each of them can describe
the experimental data in a plausible region of $\mu$.
For each of the considered $J/\psi$ LDME sets, we determine the value of $\mu$,
$\mu_{\text{fit}}$, where the respective theoretical prediction coincides with
the central value of the experimental result and list it in
Table~\ref{tab:Jpsi}.
We observe from Table~\ref{tab:Jpsi} that these $\mu_{\text{fit}}$ values all
fall into the preferred range $3.7~\text{GeV}<\mu<2\mu_{\text{FAC}}$ and thus
conclude that NRQCD factorization is compatible with experiment here.
However, we caution the reader that the $J/\psi$ LDME sets of
Refs.~\cite{Butenschoen:2011yh,Butenschoen:2012qh,Gong:2012ug,Bodwin:2015iua}
yield $\mu_{\text{fit}}$ values that are rather close to the border at 3.7~GeV,
below which the NLO prediction for $\mathcal{B}(\Upsilon\to \chi_{c0} +X)$ is
negative \cite{He:2018dho}. 

\subsection{$\Upsilon\to \chi_{cJ}+X$}

At this point, we briefly address $\Upsilon\to J/\psi+X$ via the feed-down
from $\chi_{cJ}$ mesons, {\it i.e.}\ via $\Upsilon\to\chi_{cJ}+X$ followed by
$\chi_{cJ}\to J/\psi+\gamma$.
In Ref.~\cite{He:2018dho}, we determined the $\chi_{c0}$ CO LDME to be
$\langle\mathcal{O}^{\chi_{c0}}({}^3\!S_1^{[8]})\rangle
=(4.04\pm0.47\genfrac{}{}{0pt}{2}{+0.67}{-0.34})\times10^{-3}~\text{GeV}^3$
by adopting the $\chi_{c0}$ CS LDME
$\langle\mathcal{O}^{\chi_{c0}}({}^3\!P_0^{[1]})\rangle=0.107~\text{GeV}^5$ from
the analysis of Ref.~\cite{Eichten:1995ch} with the Buchm\"uller-Type
potential~\cite{Buchmuller:1980su} and so fitting experimental data of
$\mathcal{B}(\Upsilon\to\chi_{c1}+X)$ and $\mathcal{B}(\Upsilon\to\chi_{c2}+X)$.
Using these $\chi_{c0}$ LDMEs, we find that about 20\% of
$\mathcal{B}(\Upsilon\to J/\psi+X)$ via prompt production is due to the
feed-down from $\chi_{cJ}$ mesons.

\subsection{$\Upsilon\to\eta_c+X$}

\begin{table}
  \caption{Values of $\mathcal{B}_{\text{direct}}(\Upsilon\to\eta_c+X)$ from
    Fig.~\ref{etac1} at $\mu=\mu_{\text{FAC}}$ and $\mu=\mu_{\text{fit}}$.}
\begin{ruledtabular}  
\begin{tabular}{|c|c|c|}
$J/\psi$ LDME set & $\mu=\mu_{\text{FAC}}$ & $\mu=\mu_{\text{fit}}$ \\
\hline
Butensch\"on {\it et al.}\ \cite{Butenschoen:2011yh,Butenschoen:2012qh} &
$6.2\times 10^{-4}$ & $1.0\times 10^{-3}$ \\
\hline
Gong {\it et al.}\ \cite{Gong:2012ug} & $1.9\times 10^{-3}$ &
$3.0\times 10^{-3}$ \\
\hline
Bodwin {\it et al.}\ \cite{Bodwin:2015iua} & $2.2\times 10^{-3}$ &
$3.7\times 10^{-3}$ \\
\hline
Chao {\it et al.}\ (Default) \cite{Chao:2012iv} & $1.8\times 10^{-3}$ &
$2.3\times 10^{-3}$ \\
\hline
Chao {\it et al.}\ (Set 2) \cite{Chao:2012iv} & $1.8\times 10^{-5}$ &
$2.5\times 10^{-5}$ \\
\hline
Chao {\it et al.}\ (Set 3) \cite{Han:2014jya} & $3.1\times 10^{-4}$ &
$4.3\times 10^{-4}$ \\
\end{tabular}
\end{ruledtabular}
\label{etac}
\end{table}  

\begin{figure}[!hbtp]
\centering
\includegraphics[scale=0.4]{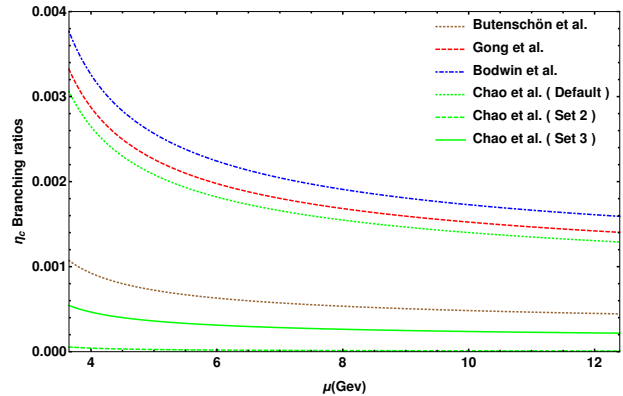}
\caption{$\mu$ dependencies of the NRQCD predictions for
  $\mathcal{B}_{\text{direct}}(\Upsilon\to\eta_c+X)$ based on the $\eta_c$ LDME
  sets obtained from Table~\ref{LDME} via Eq.~(\ref{HQSSR}).}
\label{etac1}
\end{figure}

\begin{figure*}
\centering
\includegraphics[width=0.3\linewidth]{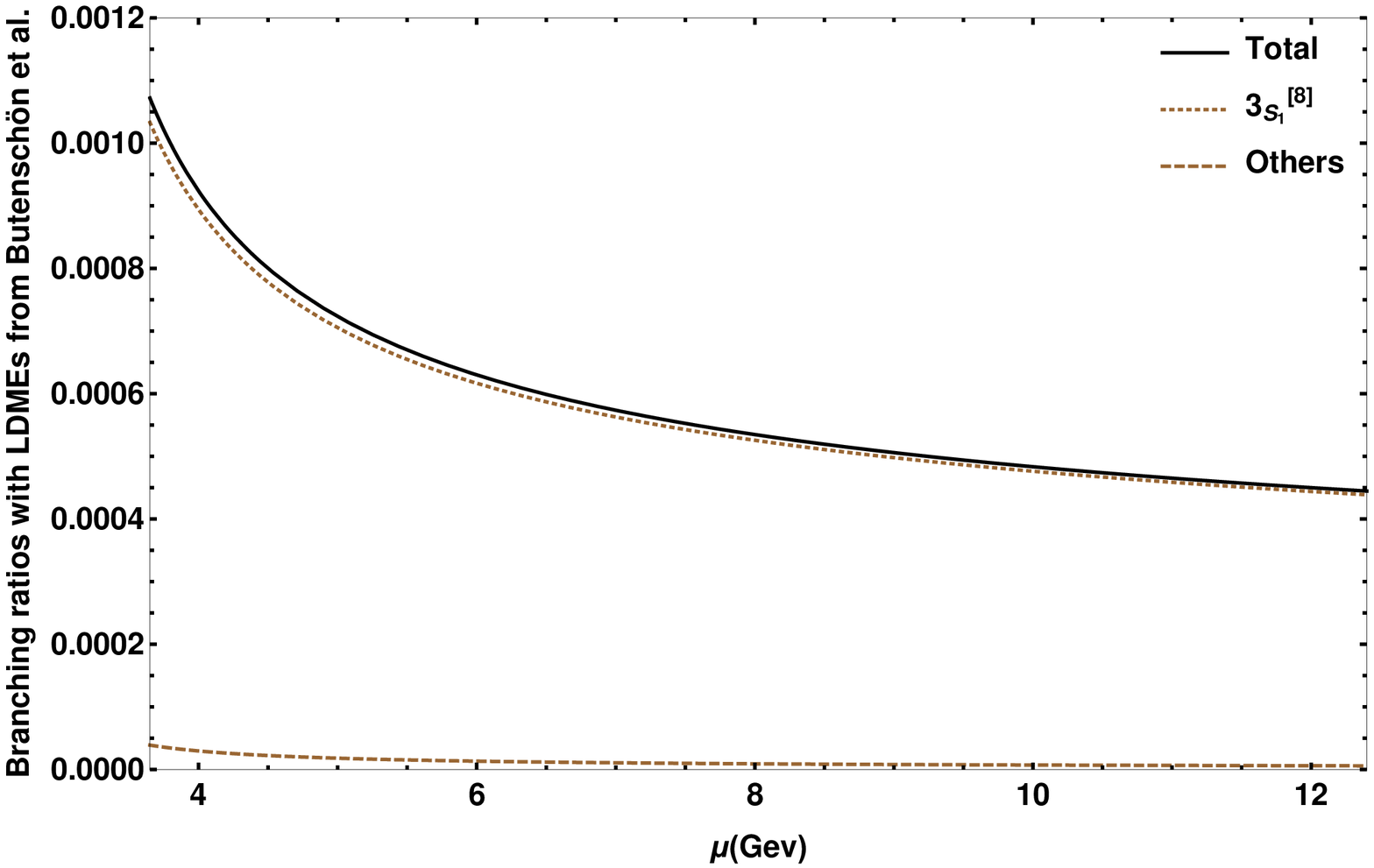}
\includegraphics[width=0.3\linewidth]{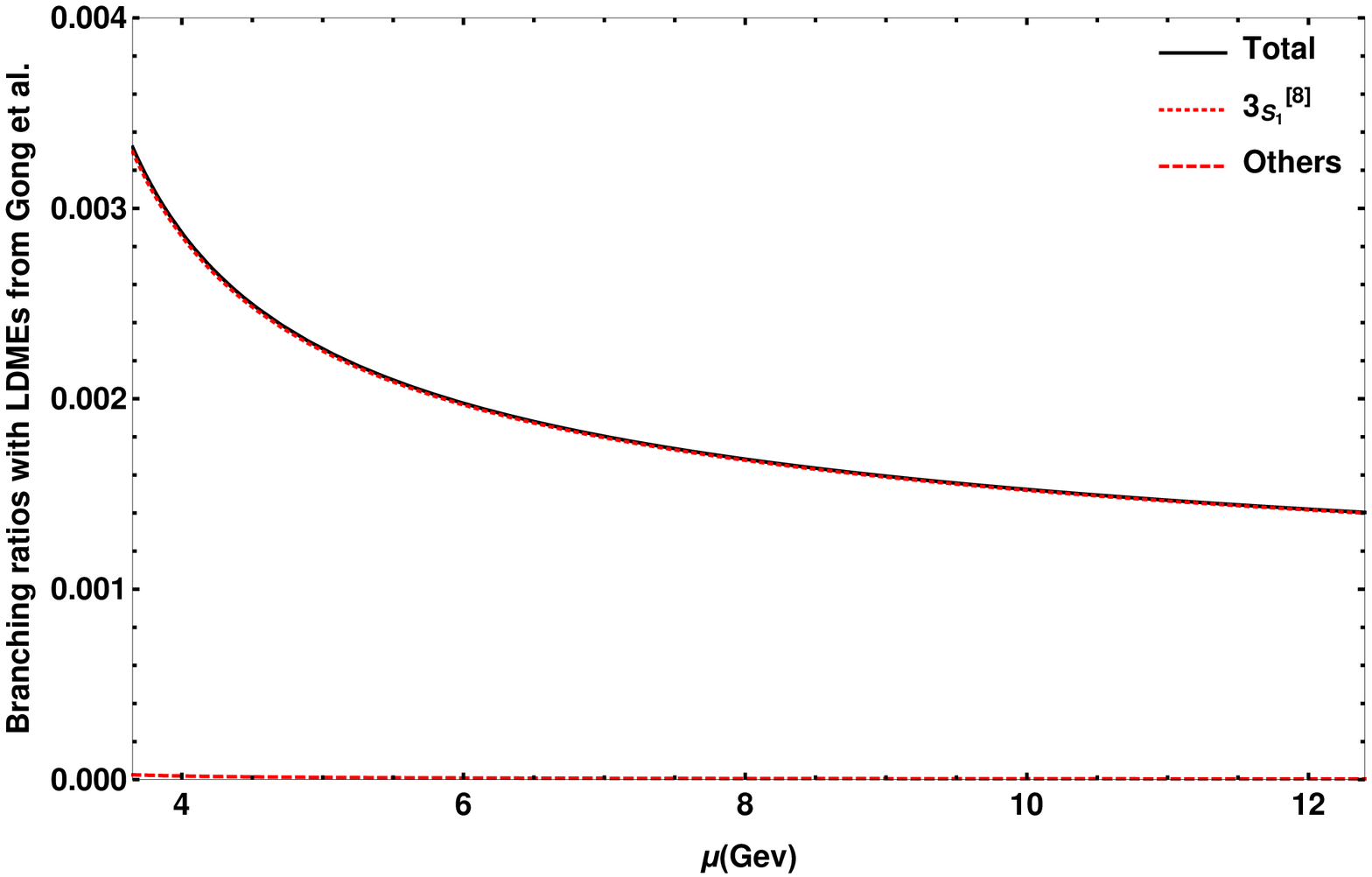}
\includegraphics[width=0.3\linewidth]{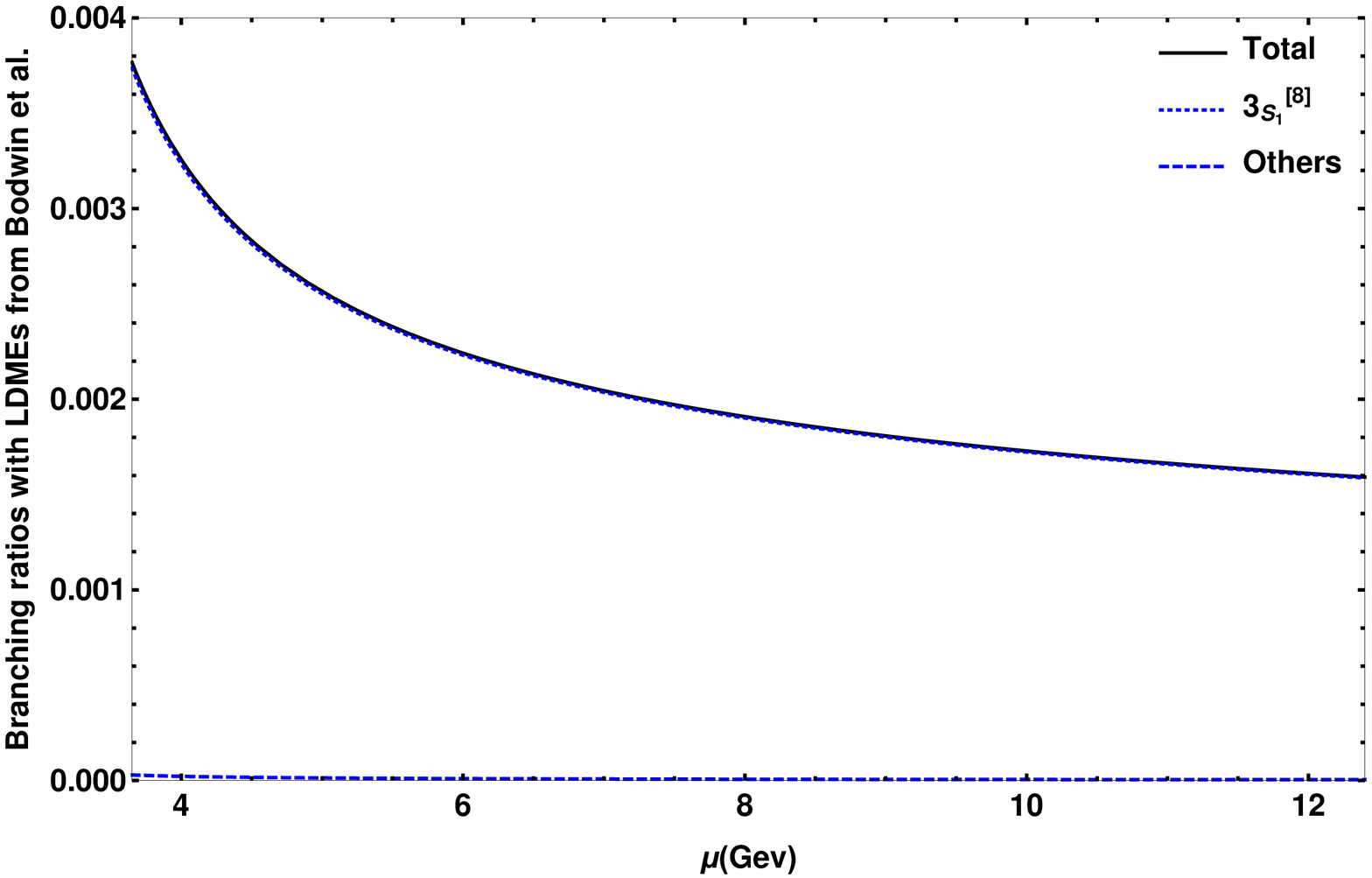}\\
\includegraphics[width=0.3\linewidth]{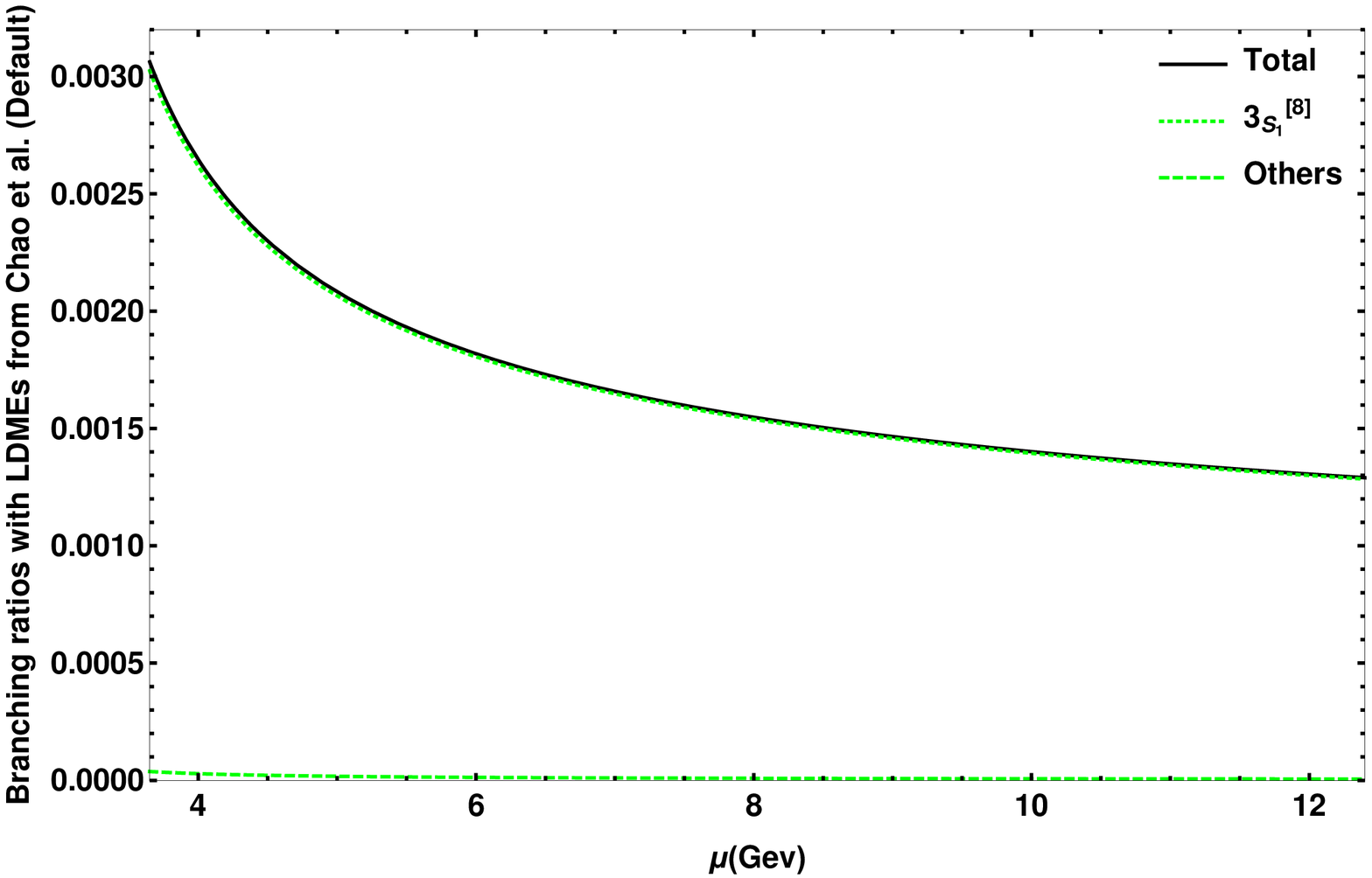}
\includegraphics[width=0.3\linewidth]{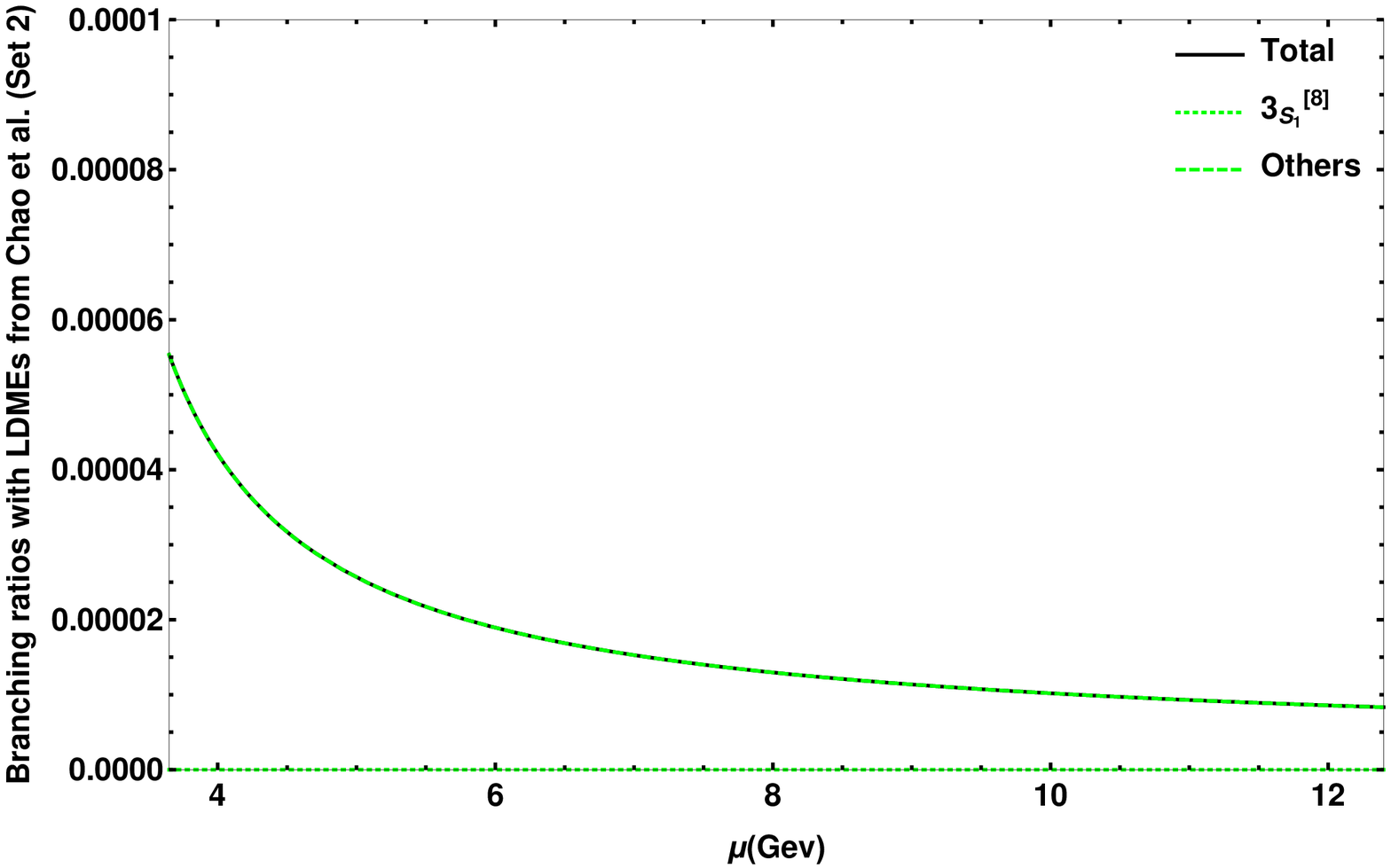}
\includegraphics[width=0.3\linewidth]{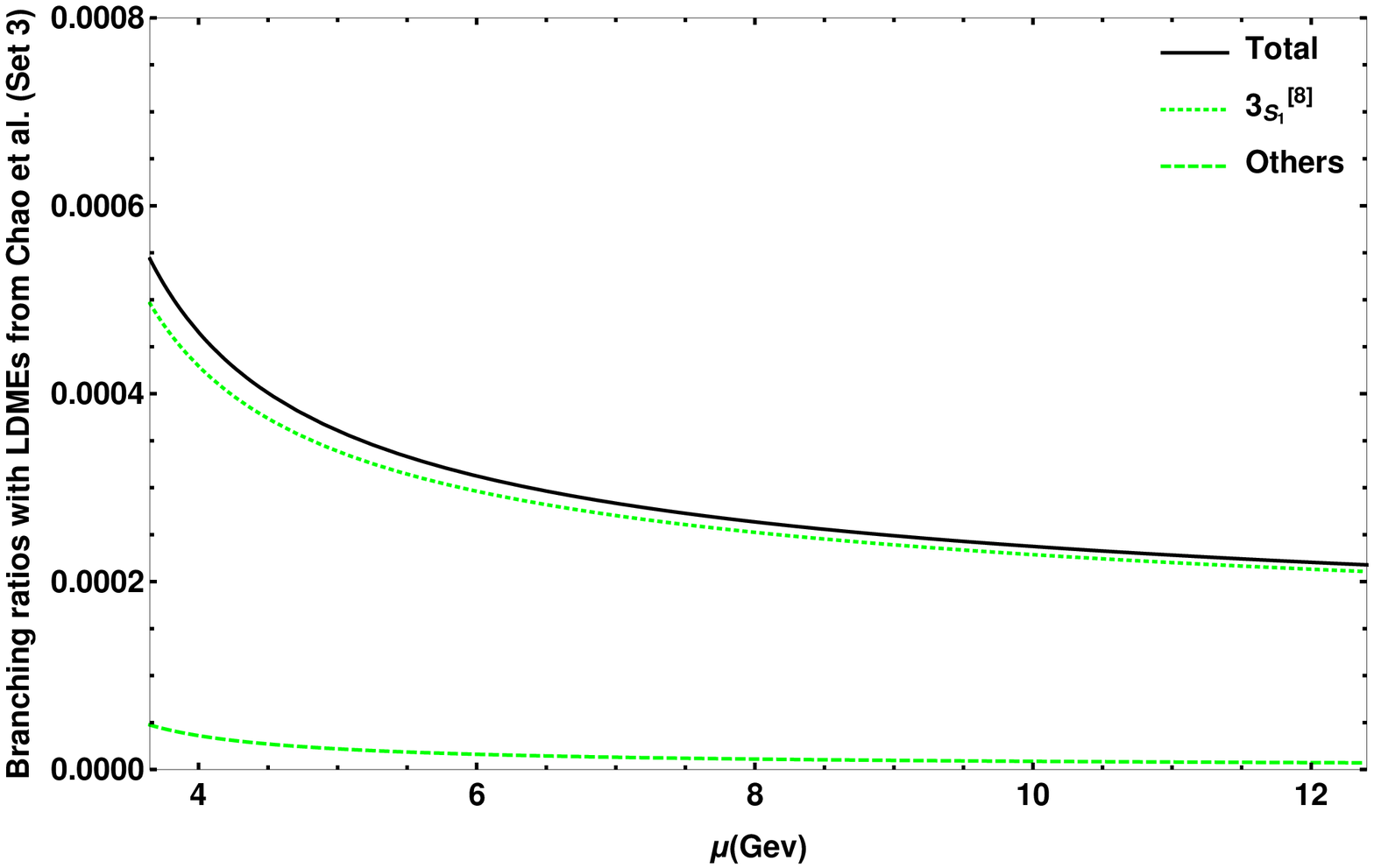}
\caption{The results in Fig.~\ref{etac1} (solid lines) are decomposed into
  their ${}^3\!S_1^{[8]}$ contributions (dotted lines) and the rest (dashed
  lines).}
\label{etac2}
\end{figure*}

We now discuss $\Upsilon\to\eta_c+X$ via prompt production.
The LDMEs of $\eta_c$ and $h_c$ production are related to those of $J/\psi$ and
$\chi_c$ production via heavy-quark spin symmetry (HQSS) at LO in $v_c^2$ as
\begin{eqnarray}\label{HQSSR}
 \langle {\cal O}^{\eta_c}({}^1\!S_0^{[1,8]}) \rangle&=& \frac{1}{3}
 \langle {\cal O}^{J/\psi}({}^3\!S_1^{[1,8]}) \rangle,\nonumber\\
 \langle {\cal O}^{\eta_c}({}^3\!S_1^{[8]}) \rangle&=&
 \langle {\cal O}^{J/\psi}({}^1\!S_0^{[8]}) \rangle,\nonumber\\
 \langle {\cal O}^{\eta_c}({}^1\!P_1^{[8]}) \rangle&=& 3
 \langle {\cal O}^{J/\psi}({}^3\!P_0^{[8]}) \rangle,\nonumber\\
 \langle {\cal O}^{h_c}({}^1\!P_1^{[1]}) \rangle&=& 3
 \langle {\cal O}^{\chi_{c0}}({}^3\!P_0^{[1]}) \rangle,\nonumber\\
 \langle {\cal O}^{h_c}({}^1\!S_0^{[8]}) \rangle&=& 3
 \langle {\cal O}^{\chi_{c0}}({}^3\!S_1^{[8]}) \rangle.
\end{eqnarray}
We now evaluate $\mathcal{B}(\Upsilon\to\eta_c+X)$ via direct production using
the SDCs in Table~\ref{SDCs} together with the $\eta_c$ LDMEs converted from
Table~\ref{LDME} via Eq.~(\ref{HQSSR}) and present the results as functions of
$\mu$ in Fig.~\ref{etac1}.
The corresponding results at $\mu=\mu_{\text{FAC}}$ are listed in
Table~\ref{etac}.
For comparison, we also present there the results at $\mu=\mu_{\text{fit}}$ for
the individual $\mu_{\text{fit}}$ values in Table~\ref{tab:Jpsi}.
We observe from Fig.~\ref{etac1} and the entries in Table~\ref{etac} for the
common scale choice $\mu=\mu_{\text{FAC}}$ that the results for the $J/\psi$
LDME sets from Refs.~\cite{Gong:2012ug,Bodwin:2015iua} and the default one
from Ref.~\cite{Chao:2012iv} are very similar, quite in contrast to the
findings in Fig.~\ref{JpsiResult} and Table~\ref{tab:Jpsi}.
This may be traced to the approximate agreement of the respective values of
$\langle\mathcal{O}^{J/\psi}({}^1\!S_0^{[8]})\rangle$ in Table~\ref{LDME} by
noticing that, similarly to the case of $\eta_c$ hadroproduction
\cite{Butenschoen:2014dra}, the bulk of the NRQCD prediction is made up by the
${}^3\!S_1^{[8]}$ channel alone.
The latter observation can be easily made from Fig.~\ref{etac2}, where the
various NRQCD predictions for $\mathcal{B}_{\text{direct}}(\Upsilon\to\eta_c+X)$
in Fig.~\ref{etac1} are decomposed into the ${}^3\!S_1^{[8]}$ contributions and
the rest.
In the special case of Set~2 of the $J/\psi$ LDMEs from
Ref.~\cite{Chao:2012iv}, where
$\langle\mathcal{O}^{J/\psi}({}^1\!S_0^{[8]})\rangle=0$, the ${}^3\!S_1^{[8]}$
channel of $\eta_c$ production is quenched leading to a particularly small NRQCD
prediction for $\mathcal{B}_{\text{direct}}(\Upsilon\to\eta_c+X)$.

In the case of $\eta_c$ inclusive hadroproduction, it was found at NLO that
the CS channel alone exhausts the cross section measured by the LHCb
Collaboration \cite{Aaij:2014bga}, while the complete NRQCD predictions
overshoot it by a few times to an order of magnitude depending on the LDME set
chosen \cite{Butenschoen:2014dra}.
It is, therefore, interesting to examine the importance of CO processes for
$\Upsilon\to\eta_c+X$.
We find that the CS contribution to
$\mathcal{B}_{\text{direct}}(\Upsilon\to\eta_c+X)$ ranges from $3.2\times10^{-5}$
to $0.5\times10^{-5}$ for $3.7~\text{GeV}<\mu<2\mu_{\text{FAC}}$ and takes the
value $1.1\times10^{-5}$ for $\mu=\mu_{\text{FAC}}$, which is about 1.6 times
smaller than the smallest NRQCD prediction for $\mu=\mu_{\text{FAC}}$ in
Table~\ref{etac}, for Set 2 of the $J/\psi$ LDMEs from Ref.~\cite{Chao:2012iv},
and more than one or even two orders of magnitude smaller than NRQCD
predictions for the other $J/\psi$ LDMEs sets.
In other words, $\Upsilon\to\eta_c+X$ is predicted to proceed even more
dominantly through CO processes than $\eta_c$ inclusive hadroproduction.
We thus conclude that the study of $\Upsilon\to\eta_c+X$ is expected to provide
a crucial test of the CO mechanism of NRQCD factorization and to place valuable
constraints on the CO LDMEs of $J/\psi$ production, in particular on
$\langle\mathcal{O}^{J/\psi}({}^1\!S_0^{[8]})\rangle$, assuming that HQSS is
preserved. 

Finally, we caution the reader that, in the case of
$\mathcal{B}_{\text{direct}}(\Upsilon\to\eta_c+X)$, the CO contributions from the
initial state may be somewhat less suppressed than expected from the velocity
scaling rules \cite{Lepage:1992tx}.
In fact, comparing the entry for the $J/\psi$ LDMEs of
Refs.~\cite{Butenschoen:2011yh,Butenschoen:2012qh} and $\mu=\mu_{\text{FAC}}$
($\mu_{\text{fit}}$) in Table~\ref{etac} with the respective entry for the hybrid
scheme in Table~\ref{tab:co}, we observe that the contribution due to the $S$
wave CO $b\bar{b}$ Fock states reaches about 26\% (16\%) of our default result.

\subsection{$\Upsilon\to h_c+X$}

\begin{figure}
\centering
\includegraphics[width=0.9\linewidth]{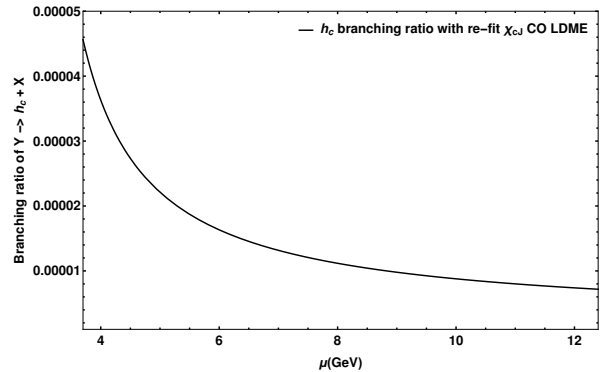}
\caption{$\mu$ dependence of NRQCD prediction for
  $\mathcal{B}(\Upsilon\to h_c+X)$ based on the $h_c$ LDME set obtained from
  Ref.~\cite{He:2018dho} via Eq.~(\ref{HQSSR}).}
\label{hcResult}
\end{figure}

Finally, we turn to $\Upsilon\to\eta_c+X$ via the feed-down from $h_c$ mesons,
{\it i.e.}\ via $\Upsilon\to h_c+X$ followed by $h_c\to\eta_c+\gamma$.
Converting the LDMEs for $\chi_{c0}$ production quoted above to those of $h_c$
production via the appropriate HQSS relations in Eq.~(\ref{HQSSR}), we obtain
$\langle\mathcal{O}^{h_{c}}({}^1\!P_1^{[1]})\rangle=0.321~\text{GeV}^5$ and
$\langle\mathcal{O}^{h_{c}}({}^1\!S_0^{[8]})\rangle
=(1.21\pm0.15{+0.20\atop-0.10})\times10^{-2}~\text{GeV}^3$.
The resulting NRQCD prediction for $\mathcal{B}(\Upsilon\to h_c+X)$ is
presented as a function of $\mu$ in Fig.~\ref{hcResult}.
At $\mu=\mu_{\text{FAC}}$, we have
$\mathcal{B}(\Upsilon\to h_c+X)=1.6\times10^{-5}$.
Taking into account the branching fraction $\mathcal{B}(h_c\to\eta_c+X)=51\%$
\cite{Tanabashi:2018oca}, the feed-down contribution is found to be
$\mathcal{B}_{h_c\text{feed down}}(\Upsilon\to \eta_c+X)=8.0\times10^{-6}$ at 
$\mu=\mu_{\text{FAC}}$.

\section{Conclusions}
\label{sec:four}

In summary, we studied $J/\psi$ and $\eta_c$ inclusive production via
$\Upsilon$ decay in the NRQCD factorization approach, including also the
feed-down from $h_c$ mesons in the latter case.
We calculated the SDCs completely through $\mathcal{O}(\alpha_s^5)$ keeping the
$b\bar{b}$ pair in the ${}^3\!S_1^{[1]}$ CS Fock state, but including
all the relevant $c\bar{c}$ Fock states, both the ${}^1\!S_0^{[1]}$,
${}^3\!S_1^{[1]}$, and ${}^1\!P_1^{[1]}$ CS ones and the ${}^1\!S_0^{[8]}$,
${}^3\!S_1^{[8]}$, ${}^1\!P_1^{[8]}$, and ${}^3\!P_J^{[8]}$ CO ones, and allowing
for all possible combinations of gluons or open $c\bar{c}$ pairs to accompany
the charmonia.

As for $\Upsilon\to J/\psi+X$ via direct production, we found that the NLO sets
of $J/\psi$ LDMEs recently determined from fits to experimental data of
inclusive production at various types of colliders or just hadron colliders all
lead to useful interpretations of the measured branching fraction, in the sense
that the renormalization scale $\mu_{\text{fit}}$ which yields agreement with
the central value is in the ballpark of the optimal scale
$\mu_{\text{FAC}}=6.2~\text{GeV}$ of the FAC principle.

In want of direct determinations, we relied on $\eta_c$ and $h_c$ LDMEs
obtained from $J/\psi$ and $\chi_{cJ}$ ones via HQSS relations.
As for $\Upsilon\to\eta_c+X$ via prompt production, we thus predicted the
branching fraction to range between $2.6\times 10^{-5}$ and $2.2\times10^{-3}$ 
at $\mu=\mu_{\text{FAC}}$, with a strong dependence on the choice of LDMEs,
in particular on the value of $\langle{\cal O}^{\eta_c}({}^3\!S_1^{[8]})\rangle$.
On the contrary, the CS contribution to
$\mathcal{B}_\text{prompt}(\Upsilon\to\eta_c+X)$ amounts to just about
$1.1\times 10^{-5}$ at $\mu=\mu_{\text{FAC}}$.
Our findings suggest that inclusive $\eta_c$ production in $\Upsilon$ decay
will provide a particularly clean laboratory to investigate the CO mechanism in
charmonium production and so to shed light on the $J/\psi$ polarization puzzle
at hadron colliders, and we propose to carry out such measurements in the
future.

\acknowledgments

We would like to thank G.~T. Bodwin and M. Butensch\"on for useful discussions.
This work was supported in part by the German Federal Ministry for Education
and Research BMBF through Grant No.\ 05H15GUCC1, by the German Research
Foundation DFG through Grant No.\ KN~365/12-1, and by the China Scholarship
Council CSC through Grant No.\ CSC-201404910576.
The work of X.-P.~W. is supported by the U.S. Department of Energy,
Division of High Energy Physics, under Contract No.\ DE-AC02-06CH11357.
The submitted manuscript has been created in part by UChicago Argonne,
LLC, Operator of Argonne National Laboratory. Argonne, a U.S.
Department of Energy Office of Science laboratory, is operated under
Contract No.\ DE-AC02-06CH11357. The U.S. Government retains for
itself, and others acting on its behalf, a paid-up nonexclusive,
irrevocable worldwide license in said article to reproduce, prepare
derivative works, distribute copies to the public, and perform publicly
and display publicly, by or on behalf of the Government.


\appendix*

\section{SDCs $\hat{\Gamma}_{mn}$ at $\mathcal{O}(\alpha_s^3)$}
\label{app:a}

\begin{table*} 
  \caption{$\mathcal{O}(\alpha_{s}^{3})$ contributions to
    $\mathcal{B}_\text{direct}(\Upsilon\to H+X)$ for
    $H=J/\psi,\eta_c,\chi_{cJ},h_c$
    due to the $S$ wave CO $b\bar{b}$ Fock states.}
\begin{ruledtabular}
  \begin{tabular}{|c|c|c|c|c|c|c|}
scheme & $J/\psi$ & $\eta_{c}$ & $\chi_{c0}$ & $\chi_{c1}$ & $\chi_{c2}$ & $h_{c}$ \\
 \hline
 hybrid & $8.3\times 10^{-6}$ & $1.6\times 10^{-4}$ & $2.0\times 10^{-5}$ & $2.2\times 10^{-5}$ & $2.1\times 10^{-5}$ & $2.0\times 10^{-7}$ \\
  \hline
  nrqcd & $5.0\times 10^{-7}$ & $9.0\times 10^{-6}$ & $8.3\times 10^{-7}$ & $1.7\times 10^{-6}$ & $1.2\times 10^{-6}$ & $1.7\times 10^{-7}$ \\
  \end{tabular}
  \end{ruledtabular}
 \label{tab:co}
\end{table*}

Here we list our analytic results for the SDCs $\hat{\Gamma}_{mn}$ of the decay
processes $\Upsilon\to H+X$ with $H=J/\psi,\eta_c,\chi_{cJ},h_c$ at
$\mathcal{O}(\alpha_s^3)$.
We have
\begin{widetext}
\begin{eqnarray}
\hat{\Gamma}(b\bar{b}({}^{1}\!{S}_{0}^{[8]})\to c\bar{c} ({}^{3}\!S_{1}^{[8]}) + g)
&=&\frac{5\pi^{2}\alpha_{s}^{3}(m_{b}^{2}-m_{c}^{2})}{72m_{b}^{4}m_{c}^{3}},
  \nonumber\\
\hat{\Gamma}(b\bar{b}({}^{3}\!{S}_{1}^{[8]})\to c\bar{c} ({}^{1}\!S_{0}^{[1]}) + g)
&=&\frac{\pi^{2}\alpha_{s}^{3}(m_{b}^{2}-m_{c}^{2})}{27m_{b}^{6}m_{c}},
  \nonumber\\
\hat{\Gamma}(b\bar{b}({}^{3}\!{S}_{1}^{[8]})\to c\bar{c} ({}^{1}\!S_{0}^{[8]}) + g)
&=&\frac{5\pi^{2}\alpha_{s}^{3}(m_{b}^{2}-m_{c}^{2})}{72m_{b}^{6}m_{c}},
  \nonumber\\
\hat{\Gamma}(b\bar{b}({}^{3}\!{S}_{1}^{[8]})\to c\bar{c} ({}^{1}\!P_{1}^{[8]}) + g)
&=&\frac{\pi^{2}\alpha_{s}^{3}(m_{b}^{2}-m_{c}^{2})}{24m_{b}^{6}m_{c}^{3}},
  \nonumber\\
\hat{\Gamma}(b\bar{b}({}^{3}\!{S}_{1}^{[8]})\to c\bar{c} ({}^{3}\!P_{0}^{[1]}) + g)
&=&\frac{\pi^{2}\alpha_{s}^{3}(m_{b}^{2}-3m_{c}^{2})^{2}}{81m_{b}^{6}m_{c}^{3}(m_{b}^{2}-m_{c}^{2})},
  \nonumber\\
\hat{\Gamma}(b\bar{b}({}^{3}\!{S}_{1}^{[8]})\to c\bar{c} ({}^{3}\!P_{1}^{[1]}) + g)
&=&\frac{2\pi^{2}\alpha_{s}^{3}(m_{b}^{2}+m_{c}^{2})}{81m_{b}^{4}m_{c}^{3}(m_{b}^{2}-m_{c}^{2})},
  \nonumber\\
\hat{\Gamma}(b\bar{b}({}^{3}\!{S}_{1}^{[8]})\to c\bar{c} ({}^{3}\!P_{2}^{[1]}) + g)
&=&\frac{2\pi^{2}\alpha_{s}^{3}(m_{b}^{4}+3m_{b}^{2}m_{c}^{2}+6m_{c}^{4})}{405m_{b}^{6}m_{c}^{3}(m_{b}^{2}-m_{c}^{2})},
  \nonumber\\
\sum_{J=0}^2\hat{\Gamma}(b\bar{b}({}^{3}\!{S}_{1}^{[8]})\to c\bar{c} ({}^{3}\!P_{J}^{[8]}) + g)
&=&\frac{5\pi^{2}\alpha_{s}^{3}(3m_{b}^{4}+2m_{b}^{2}m_{c}^{2}+7m_{c}^{4})}{72m_{b}^{6}m_{c}^{3}(m_{b}^{2}-m_{c}^{2})},
\nonumber\\
\sum_{J=0}^2\hat{\Gamma}(b\bar{b}({}^{3}\!{P}_{J}^{[8]})\to c\bar{c} ({}^{3}\!S_{1}^{[8]}) + g)
&=&\frac{5\pi^{2}\alpha_{s}^{3}(7m_{b}^{4}+2m_{b}^{2}m_{c}^{2}+3m_{c}^{4})}{72m_{b}^{6}m_{c}^{3}(m_{b}^{2}-m_{c}^{2})}.
\end{eqnarray}
\end{widetext}
We note that, incidentally,
$\hat{\Gamma}(b\bar{b}({}^{3}\!{S}_{1}^{[8]})\to c\bar{c}({}^{3}\!S_{1}^{[8]})+g)=0$
at $\mathcal{O}(\alpha_s^3)$.
Adopting the $J/\psi$ LDMEs from
Refs.~\cite{Butenschoen:2011yh,Butenschoen:2012qh} and the $\chi_{cJ}$ ones
from Ref.~\cite{He:2018dho} and deriving from them the $\eta_c$ and $h_c$ ones
via the HQSS relations in Eq.~(\ref{HQSSR}), we find that the results for the
ratios
$\langle\Upsilon|\mathcal{O}({}^1S_0^{[8]})|\Upsilon\rangle
/\langle\Upsilon|\mathcal{O}({}^3S_1^{[1]})|\Upsilon\rangle$ and
$\langle\Upsilon|\mathcal{O}({}^3S_8^{[8]})|\Upsilon\rangle
/\langle\Upsilon|\mathcal{O}({}^3S_1^{[1]})|\Upsilon\rangle$
determined in Ref.~\cite{Bodwin:2005gg}, namely $2.414(3)\times10^{-3}$ and
$8.1(6)\times10^{-5}$ in the hybrid scheme and $9.0(1)\times10^{-5}$ and
$6.9(5)\times10^{-5}$ in the nrqcd scheme, yield the $\mathcal{O}(\alpha_s^3)$
contributions to $\mathcal{B}_\text{direct}(\Upsilon\to H+X)$ specified in
Table~\ref{tab:co}.
These results are $\mu$ independent thanks to the normalization in
Eq.~(\ref{eq:norm}), which, for consistency, is evaluated here at LO.
They still need to be supplemented by the contributions proportional to
$\langle\Upsilon|\mathcal{O}({}^3P_0^{[8]})|\Upsilon\rangle$, which is still
unknown.


\begin{thebibliography}{99}

\bibitem{Caswell:1985ui} 
  W.~E.~Caswell and G.~P.~Lepage,
  Effective lagrangians for Bound State Problems in QED, QCD, and Other Field Theories,
  Phys.\ Lett.\  {\bf 167B}, 437 (1986).

\bibitem{Bodwin:1994jh} 
  G.~T.~Bodwin, E.~Braaten, and G.~P.~Lepage,
  Rigorous QCD analysis of inclusive annihilation and production of heavy quarkonium,
  Phys.\ Rev.\ D {\bf 51}, 1125 (1995);
  {\bf 55}, 5853(E) (1997)
   [hep-ph/9407339].

\bibitem{Lepage:1992tx} 
  G.~P.~Lepage, L.~Magnea, C.~Nakhleh, U.~Magnea, and K.~Hornbostel,
  Improved nonrelativistic QCD for heavy-quark physics,
  Phys.\ Rev.\ D {\bf 46}, 4052 (1992)
  [hep-lat/9205007].

\bibitem{Brambilla:2010cs} 
  N.~Brambilla {\it et al.},
  Heavy quarkonium: progress, puzzles, and opportunities,
  Eur.\ Phys.\ J.\ C {\bf 71}, 1534 (2011)
  [arXiv:1010.5827 [hep-ph]].

\bibitem{Brambilla:2014jmp} 
  N.~Brambilla {\it et al.},
  QCD and strongly coupled gauge theories: challenges and perspectives,
  Eur.\ Phys.\ J.\ C {\bf 74}, 
  2981 (2014)
  [arXiv:1404.3723 [hep-ph]].
 
\bibitem{Lansberg:2019adr} 
  J.-P.~Lansberg,
  New Observables in Inclusive Production of Quarkonia,
  arXiv:1903.09185 [hep-ph].

\bibitem{Zhang:2009ym} 
  Y.-J.~Zhang, Y.-Q.~Ma, K.~Wang, and K.-T.~Chao,
  QCD radiative correction to color-octet $J/\psi$ inclusive production at $B$ Factories,
  Phys.\ Rev.\ D {\bf 81}, 034015 (2010)
  [arXiv:0911.2166 [hep-ph]].

\bibitem{Ma:2008gq} 
  Y.-Q.~Ma, Y.-J.~Zhang, and K.-T.~Chao,
  QCD Corrections to $e^+ e^- \to J/\psi+g g$ at $B$ Factories,
  Phys.\ Rev.\ Lett.\  {\bf 102}, 162002 (2009)
  [arXiv:0812.5106 [hep-ph]].

\bibitem{Gong:2009kp} 
  B.~Gong and J.-X.~Wang,
  Next-to-Leading-Order QCD Corrections to $e^+ e^- \to J/\psi gg$ at the $B$ Factories,
  Phys.\ Rev.\ Lett.\  {\bf 102}, 162003 (2009)
  [arXiv:0901.0117 [hep-ph]].

\bibitem{Klasen:2004tz} 
  M.~Klasen, B.~A.~Kniehl, L.~N.~Mihaila, and M.~Steinhauser,
  $J/\psi$ plus jet associated production in two-photon collisions at next-to-leading order,
  Nucl.\ Phys.\ B {\bf 713}, 487 (2005)
  [hep-ph/0407014].

\bibitem{Klasen:2004az} 
  M.~Klasen, B.~A.~Kniehl, L.~N.~Mihaila, and M.~Steinhauser,
  $J/\psi$ plus prompt-photon associated production in two-photon collisions at next-to-leading order,
  Phys.\ Rev.\ D {\bf 71}, 014016 (2005)
  [hep-ph/0408280].

\bibitem{Butenschoen:2011yh} 
  M.~Butenschoen and B.~A.~Kniehl,
  World data of $J/\psi$ production consolidate nonrelativistic QCD factorization at next-to-leading order,
  Phys.\ Rev.\ D {\bf 84}, 051501(R) (2011)
  [arXiv:1105.0820 [hep-ph]].

\bibitem{Butenschoen:2009zy} 
  M.~Butensch\"on and B.~A.~Kniehl,
  Complete Next-to-Leading-Order Corrections to $J/\psi$ Photoproduction in Nonrelativistic Quantum Chromodynamics,
  Phys.\ Rev.\ Lett.\  {\bf 104}, 072001 (2010)
  [arXiv:0909.2798 [hep-ph]].

\bibitem{Butenschoen:2011ks} 
  M.~Butenschoen and B.~A.~Kniehl,
  Probing Nonrelativistic QCD Factorization in Polarized $J/\psi$ Photoproduction at Next-to-Leading Order,
  Phys.\ Rev.\ Lett.\  {\bf 107}, 232001 (2011)
  [arXiv:1109.1476 [hep-ph]].
  
\bibitem{Ma:2010yw} 
  Y.-Q.~Ma, K.~Wang, and K.-T.~Chao,
  $J/\psi (\psi^\prime)$ Production at the Tevatron and LHC at ${\cal O}(\alpha_s^4v^4)$ in Nonrelativistic QCD,
  Phys.\ Rev.\ Lett.\  {\bf 106}, 042002 (2011)
  [arXiv:1009.3655 [hep-ph]]

\bibitem{Butenschoen:2010rq} 
  M.~Butensch\"on and B.~A.~Kniehl,
  Reconciling $J/\psi$ Production at HERA, RHIC, Tevatron, and LHC with Nonrelativistic QCD Factorization at Next-to-Leading Order,
  Phys.\ Rev.\ Lett.\  {\bf 106}, 022003 (2011).
  [arXiv:1009.5662 [hep-ph]].

\bibitem{Butenschoen:2012px} 
  M.~Butenschoen and B.~A.~Kniehl,
  $J/\psi$ Polarization at the Tevatron and the LHC: Nonrelativistic-QCD Factorization at the Crossroads,
  Phys.\ Rev.\ Lett.\  {\bf 108}, 172002 (2012)
  [arXiv:1201.1872 [hep-ph]].

\bibitem{Chao:2012iv} 
  K.-T.~Chao, Y.-Q.~Ma, H.-S.~Shao, K.~Wang, and Y.-J.~Zhang,
  $J/\psi$ Polarization at Hadron Colliders in Nonrelativistic QCD,
  Phys.\ Rev.\ Lett.\  {\bf 108}, 242004 (2012)
  [arXiv:1201.2675 [hep-ph]].

\bibitem{Gong:2012ug} 
  B.~Gong, L.-P.~Wan, J.-X.~Wang, and H.-F.~Zhang,
  Polarization for Prompt $J/\psi$ and $\psi(2s)$ Production at the Tevatron and LHC,
  Phys.\ Rev.\ Lett.\  {\bf 110}, 042002 (2013)
  [arXiv:1205.6682 [hep-ph]].

\bibitem{Shao:2014yta} 
  H.-S.~Shao, H.~Han, Y.-Q.~Ma, C.~Meng, Y.-J.~Zhang, and K.-T.~Chao,
  Yields and polarizations of prompt $J/\psi$ and $\psi(2S)$ production in hadronic collisions,
  J. High Energy Phys.\ 05 (2015) 103
  [arXiv:1411.3300 [hep-ph]].

\bibitem{Butenschoen:2014dra} 
  M.~Butenschoen, Z.-G.~He, and B.~A.~Kniehl,
  $\eta_c$ Production at the LHC Challenges Nonrelativistic QCD Factorization,
  Phys.\ Rev.\ Lett.\  {\bf 114}, 
  092004 (2015)
  [arXiv:1411.5287 [hep-ph]].

\bibitem{Aaij:2014bga} 
  R.~Aaij {\it et al.}\ (LHCb Collaboration),
  Measurement of the $\eta_c (1S)$ production cross-section in proton-proton collisions 
  via the decay $\eta_c (1S) \rightarrow p \bar{p}$,
  Eur.\ Phys.\ J.\ C {\bf 75}, 
  311 (2015)
  [arXiv:1409.3612 [hep-ex]].

\bibitem{He:2018dho} 
  Z.-G.~He, B.~A.~Kniehl, and X.-P.~Wang,
  Inclusive $\chi_{cJ}$ production in $\Upsilon$ decay at $\mathcal{O}(\alpha_s^5)$ in NRQCD factorization,
  Phys.\ Rev.\ D {\bf 98}, 
  074005 (2018)
  [arXiv:1809.00612 [hep-ph]].

\bibitem{Fulton:1988ug} 
  R.~Fulton {\it et al.}\ (CLEO Collaboration),
  First observation of inclusive $\psi$ production in $\Upsilon$ decays,
  Phys.\ Lett.\ B {\bf 224}, 445 (1989).

\bibitem{Maschmann:1989ai} 
  W.~Maschmann {\it et al.}\ (Crystal Ball Collaboration),
  Inclusive $J/\psi$ production in decays of $B$ mesons,
  Z.\ Phys.\ C {\bf 46}, 555 (1990).

\bibitem{Albrecht:1992ap} 
  H.~Albrecht {\it et al.}\ (ARGUS Collaboration),
  Search for charm production in direct decays of the $\Upsilon (1S)$ resonance,
  Z.\ Phys.\ C {\bf 55}, 25 (1992).

\bibitem{Briere:2004ug} 
  R.~A.~Briere {\it et al.}\ (CLEO Collaboration),
  New measurements of $\Upsilon(1S)$ decays to charmonium final states,
  Phys.\ Rev.\ D {\bf 70}, 072001 (2004)
  [hep-ex/0407030].

\bibitem{Shen:2016yzg} 
  C.~P.~Shen {\it et al.}\ (Belle Collaboration),
  Search for $XYZ$ states in $\Upsilon(1S)$ inclusive decays,
  Phys.\ Rev.\ D {\bf 93}, 
  112013 (2016)
  [arXiv:1605.00990 [hep-ex]].

\bibitem{Tanabashi:2018oca} 
  M.~Tanabashi {\it et al.}\ (Particle Data Group),
  Review of Particle Physics,
  Phys.\ Rev.\ D {\bf 98}, 
  030001 (2018).

\bibitem{Cheung:1996mh} 
  K.~Cheung, W.-Y.~Keung, and T.~C.~Yuan,
  Color-octet $J/\psi$ production in the $\Upsilon$ decay,
  Phys.\ Rev.\ D {\bf 54}, 929 (1996)
  [hep-ph/9602423].

\bibitem{Napsuciale:1997bz} 
  M.~Napsuciale,
  Inclusive $J/\psi$ production in $\Upsilon$ decay via color octet mechanisms,
  Phys.\ Rev.\ D {\bf 57}, 5711 (1998)
  [hep-ph/9710488].

\bibitem{He:2009by} 
  Z.-G.~He and J.-X.~Wang,
  Inclusive $J/\psi$ production in $\Upsilon$ decay via color-singlet mechanism,
  Phys.\ Rev.\ D {\bf 81}, 054030 (2010)
  [arXiv:0911.0139 [hep-ph]].
   
\bibitem{He:2010cb} 
  Z.-G.~He and J.-X.~Wang,
  Color-singlet $J/\psi$ production at $\mathcal{O}(\alpha_s^6)$ in $\Upsilon$ decay,
  Phys.\ Rev.\ D {\bf 82}, 094033 (2010)
  [arXiv:1009.1563 [hep-ph]].

\bibitem{Bodwin:2005gg} 
  G.~T.~Bodwin, J.~Lee, and D.~K.~Sinclair,
  Spin correlations and velocity-scaling in color-octet NRQCD matrix elements,
  Phys.\ Rev.\ D {\bf 72}, 014009 (2005)
  [hep-lat/0503032].
  
\bibitem{Hahn:2000kx} 
  T.~Hahn,
  Generating Feynman diagrams and amplitudes with {\it FeynArts} 3,
  Comput.\ Phys.\ Commun.\  {\bf 140}, 418 (2001)
  [hep-ph/0012260].

\bibitem{Mertig:1990an} 
  R.~Mertig, M.~B\"ohm, and A.~Denner,
  Feyn Calc --- Computer-algebraic calculation of Feynman amplitudes,
  Comput.\ Phys.\ Commun.\  {\bf 64}, 345 (1991).

\bibitem{Kuipers:2012rf} 
  J.~Kuipers, T.~Ueda, J.~A.~M.~Vermaseren, and J.~Vollinga,
  FORM version 4.0,
  Comput.\ Phys.\ Commun.\  {\bf 184}, 1453 (2013)
  [arXiv:1203.6543 [cs.SC]].

\bibitem{Feng:2012iq} 
  F.~Feng,
  {\tt \$Apart}: A generalized {\sc Mathematica} {\tt Apart} function,
  Comput.\ Phys.\ Commun.\  {\bf 183}, 2158 (2012)
  [arXiv:1204.2314 [hep-ph]].

\bibitem{Smirnov:2008iw} 
  A.~V.~Smirnov,
  Algorithm FIRE -- Feynman Integral REduction,
  JHEP {\bf 0810}, 107 (2008)
  [arXiv:0807.3243 [hep-ph]].

\bibitem{Carrazza:2016gav} 
  S.~Carrazza, R.~K.~Ellis, and G.~Zanderighi,
  {\tt QCDLoop}: a comprehensive framework for one-loop scalar integrals,
  Comput.\ Phys.\ Commun.\  {\bf 209}, 134 (2016)
  [arXiv:1605.03181 [hep-ph]].

\bibitem{Hahn:2004fe} 
  T.~Hahn,
  {\sc Cuba}: A Library for multidimensional numerical integration,
  Comput.\ Phys.\ Commun.\  {\bf 168}, 78 (2005)
  [hep-ph/0404043].

\bibitem{Korner:1982vg}
  J.~G.~K\"orner, J.~H.~K\"uhn, M.~Krammer, and H.~Schneider,
  Zweig-forbidden radiative orthoquarkonium decays in perturbative QCD,
  Nucl.\ Phys.\ B {\bf 229} (1983) 115.

\bibitem{Harris:2001sx} 
  B.~W.~Harris and J.~F.~Owens,
  The two cutoff phase space slicing method,
  Phys.\ Rev.\ D {\bf 65}, 094032 (2002)
  [hep-ph/0102128].

 \bibitem{Butenschon:2009zza} 
   M.~Butensch\"on,
   Photoproduction of the $J/\psi$ meson at HERA at next-to-leading order within the frmework of nonrelativistic QCD,
   DESY-THESIS-2009-021.
   
\bibitem{Li:1999ar} 
  S.-y.~Li, Q.-b.~Xie, and Q.~Wang,
  Contribution of colour-singlet process $\Upsilon \to J / \psi + c \bar{c} g$ to $\Upsilon \to J / \psi + X$,
  Phys.\ Lett.\ B {\bf 482}, 65 (2000)
  [hep-ph/9912328].

\bibitem{Grunberg:1980ja} 
  G.~Grunberg,
  Renormalization group improved perturbative QCD,
  Phys.\ Lett.\  {\bf 95B}, 70 (1980);
  {\bf 110B}, 501(E) (1982).

\bibitem{Mackenzie:1981sf} 
  P.~B.~Mackenzie and G.~P.~Lepage,
  Quantum Chromodynamic Corrections to the Gluonic Width of the $\Upsilon$ Meson,
  Phys.\ Rev.\ Lett.\  {\bf 47}, 1244 (1981).

\bibitem{Han:2014jya}
  H.~Han, Y.-Q.~Ma, C.~Meng, H.-S.~Shao, and K.-T.~Chao,
  $\eta_c$ Production at LHC and Implications for the Understanding of $J/\psi$ Production,
  Phys.\ Rev.\ Lett.\  {\bf 114}, 092005 (2015)
  [arXiv:1411.7350 [hep-ph]].

 \bibitem{Butenschoen:2012qh}
   M.~Butenschoen and B.~A.~Kniehl,
   $J/\psi$ production in NRQCD: A global analysis of yield and polarization,
   Nucl.\ Phys.\ Proc.\ Suppl.\  {\bf 222-224}, 151 (2012)
  [arXiv:1201.3862 [hep-ph]].

\bibitem{Bodwin:2015iua} 
  G.~T.~Bodwin, K.-T.~Chao, H.~S.~Chung, U-R.~Kim, J.~Lee, and Y.-Q.~Ma,
  Fragmentation contributions to hadroproduction of prompt$J/\psi$, $\chi_{cJ}$, and $\psi(2S)$ states,
  Phys.\ Rev.\ D {\bf 93}, 034041 (2016)
  [arXiv:1509.07904 [hep-ph]].

\bibitem{Eichten:1995ch} 
  E.~J.~Eichten and C.~Quigg,
  Quarkonium wave functions at the origin,
  Phys.\ Rev.\ D {\bf 52}, 1726 (1995)
  [hep-ph/9503356].

\bibitem{Buchmuller:1980su} 
  W.~Buchm\"uller and S.-H.~H.~Tye,
  Quarkonia and quantum chromodynamics,
  Phys.\ Rev.\ D {\bf 24}, 132 (1981).
  
\end{thebibliography}
\end{document}